\documentclass[]{aastex631}
\graphicspath{{./}{figures/}}

\begin{document}

\title{FAST search for circumstellar atomic hydrogen--I: the young planetary nebula IC\,4997}

\correspondingauthor{Yong Zhang}
\email{zhangyong5@mail.sysu.edu.cn}

\author[0000-0002-2762-6519]{Xu-Jia Ouyang}
\affiliation{School of Physics and Astronomy, Sun Yat-sen University, 2 Daxue Road, Tangjia, Zhuhai, Guangdong Province,  China}

\author[0000-0002-1086-7922]{Yong Zhang}
\affiliation{School of Physics and Astronomy, Sun Yat-sen University, 2 Daxue Road, Tangjia, Zhuhai, Guangdong Province,  China}
\affiliation{CSST Science Center for the Guangdong-Hongkong-Macau Greater Bay Area, Sun Yat-Sen University, Guangdong Province, China}
\affiliation{Laboratory for Space Research, The University of Hong Kong, Hong Kong, China}

\author[0000-0002-3171-5469]{Albert Zijlstra}
\affiliation{Department of Physics and Astronomy, The University of Manchester, Manchester M13 9PL, UK}
\affiliation{Laboratory for Space Research, The University of Hong Kong, Hong Kong, China}

\author[0000-0002-4428-3183]{Chuan-Peng Zhang}
\affiliation{National Astronomical Observatories, Chinese Academy of Sciences, Beijing 100101, China}
\affiliation{CAS Key Laboratory of FAST, National Astronomical Observatories, Chinese Academy of Sciences, Beijing 100101, China}

\author[0000-0003-3324-9462]{Jun-ichi Nakashima}
\affiliation{School of Physics and Astronomy, Sun Yat-sen University, 2 Daxue Road, Tangjia, Zhuhai, Guangdong Province,  China}
\affiliation{CSST Science Center for the Guangdong-Hongkong-Macau Greater Bay Area, Sun Yat-Sen University, Guangdong Province, China}
                
\author[0000-0002-2062-0173]{Quentin A Parker}
\affiliation{Department of Physics, The University of Hong Kong, Chong Yuet Ming Physics Building Pokfulam Road, Hong Kong, China}
\affiliation{Laboratory for Space Research, The University of Hong Kong, Hong Kong, China}
\affiliation{CSST Science Center for the Guangdong-Hongkong-Macau Greater Bay Area, Sun Yat-Sen University, Guangdong Province, China}



\begin{abstract}

Using the Five-hundred-meter Aperture Spherical radio Telescope (FAST) in Guizhou, China, we detect the 
21\,cm neutral atomic hydrogen absorption in the young planetary nebula IC\,4997. The absorption arises from a shell also associated with  \ion{Na}{1} D lines.
The \ion{H}{1} shell has a mass of
$1.46\times10^{-2}$\,M$_\sun$ and a dynamic age of 990\,yr.
The column density of \ion{H}{1} is estimated to be
$7.1\times10^{20}$\,cm$^{-2}$, which can be well explained in terms of a photodissociation region around the ionized nebula, limited by self shielding of H$_2$.
We find that the atomic-to-ionized hydrogen ratio is 0.6, suggesting that \ion{H}{1}  substantially contributes to overall nebular mass. 
\end{abstract}



\section{Introduction} \label{sec:intro}

Planetary nebulae (PNe) are the ionized, ejected remnants of the stellar winds from stars of low to intermediate mass (0.8 to 8\,M$_\sun$)  at the end of their lives.
They serve as important tools for understanding stellar evolution and the life cycle of dispersed materials in a Galaxy.
However, observations show that the total mass of the central star and ionized nebula of up to 1.5\,M$_{\sun}$ \citep{2012A&A...541A.112K} is significantly lower than the theoretical upper limit, indicating not all of the mass ejected by the star is seen in the PN.
Indeed, the total mass usually estimated for the main ionised PN shell itself is only $\sim 0.1$\,M$_{\sun}$\citep[see, e.g.,][]{2003ApJ...597..298V}. The connection between the birth mass of PN central stars and the mass left when they die (the initial-final
mass relation) is thus poorly known,  \citep[leading to the so-called 'PN missing mass problem', see][]{1994PASP..106..344K}. Such gaps in our knowledge on the abundant PN progenitor stars, which dominate the recycling of stellar material, severely impact
our ability to model the evolution and chemical enrichment of any galaxy. The asymptotic giant branch (AGB)  halos \citep[see, e.g.,][]{2003MNRAS.340..417C} are thought to be the reservoirs for much of this missing mass.
Combining the ionized gas masses derived from radio observations with the calculated molecular masses, \citet{1996A&A...315..284H} obtained nebular masses ranging from 10$^{-3}$ to 1 M$_\sun$ and found that the median value of molecular mass is 0.031 M$_\sun$ in 44 PNe.
Many PNe are ionization bounded, where part of the matter lost by the stellar wind during the pre-PN stage has not yet been ionized by the central star. A substantial part of their envelope may remain neutral \citep[see,  e.g.,][for a review]{2022PASP..134b2001K}. These outer regions of PNe show dust emission, but the abundance of refractory elements limits the dust mass to no more than 1\% of the total mass of the nebula. Infrared observations  suggest a lower dust-to-gas mass   \citep{2007MNRAS.381..117P,2020MNRAS.491..758A}, and hence most of the missing PN mass  cannot be in dust grains.  More recently, studies of molecules and ions in PNe which have evolved from common envelopes of binary stars have also shown that the missing mass is not present in these forms in these objects \citep{2021arXiv210802199S,2022A&A...658A..17S}.

However, there remains a vital circumstellar component that has not been sufficiently investigated and that could significantly contribute to the overall mass-loss budget: atomic hydrogen.

The presence of atomic hydrogen can be indicated 
through the spin-flip transition at 21\,cm wavelength in the radio regime.
Unfortunately, the detection of circumstellar atomic hydrogen
is hampered by the omnipresence of 21\,cm emission from the ambient interstellar medium.
\citet{1982Natur.299..323R} were the first to detect \ion{H}{1} absorption in the young PN NGC\,6302 using the Very Large Array after earlier unsuccessful attempts \citep{1970ApJ...160..363T,1980ApJ...241.1014Z}, and  soon thereafter, \citet[][A86 hereafter]{1986ApJ...305L..85A}  detected circumstellar \ion{H}{1} absorption towards IC\,4997 using the Arecibo telescope. Subsequent investigations confirmed  the presence of circumstellar \ion{H}{1} absorption
in five PNe and circumstellar \ion{H}{1} emission in one PN  \citep{1990ApJ...351..515T}.
To date, definite circumstellar \ion{H}{1} absorption has been detected in nine PNe \citep{1982Natur.299..323R,1986ApJ...305L..85A,1987ApJ...314..572S,1990ApJ...351..515T}, and five PNe have been detected with 
circumstellar \ion{H}{1} emission
 \citep{1987A&A...176L...5T,1990ApJ...351..515T,2000RMxAA..36...51R,2002ApJ...574..179R,2006AJ....132.2566G}.

The  number of PNe with  circumstellar \ion{H}{1} detections is extreme 
small, compared to the total number of verified Galactic PNe \citep[$\sim$3800; e.g.][]{1992secg.book.....A,2001A&A...378..843K,2005MNRAS.362..753D,2006MNRAS.373...79P,2008MNRAS.384..525M,2016JPhCS.728c2008P}. 
To improve our knowledge on the mass and dynamics of the atomic gas budget in PNe,  we have begun a program to search for \ion{H}{1} feature arising from PNe using
the Five-hundred-meter Aperture Spherical radio Telescope \citep[FAST;][]{2011IJMPD..20..989N}, the most sensitive single dish telescope in L-band available today \citep{2020RAA....20...64J}.
In this paper, we report our first results for IC\,4997 ($l=58^{\circ}.33,b=-10^{\circ}.98$),
a compact, young, rapidly evolving, well-studied, high-surface brightness PN with a prominent central star.



\section{Observation and data reduction} \label{sec:ob}

\begin{figure*}
\gridline{\fig{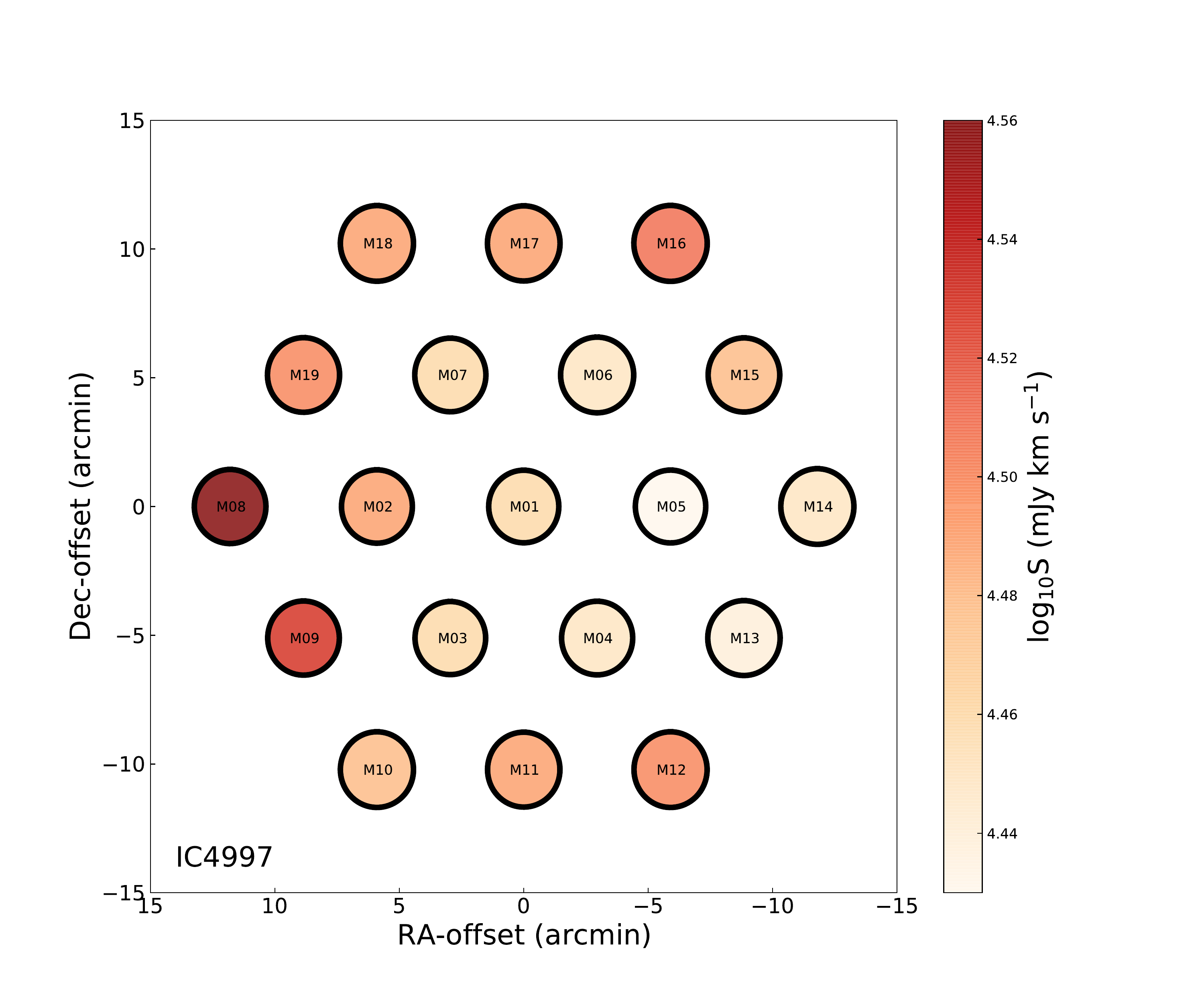}{0.6\textwidth}{(a)}
          }
\gridline{\fig{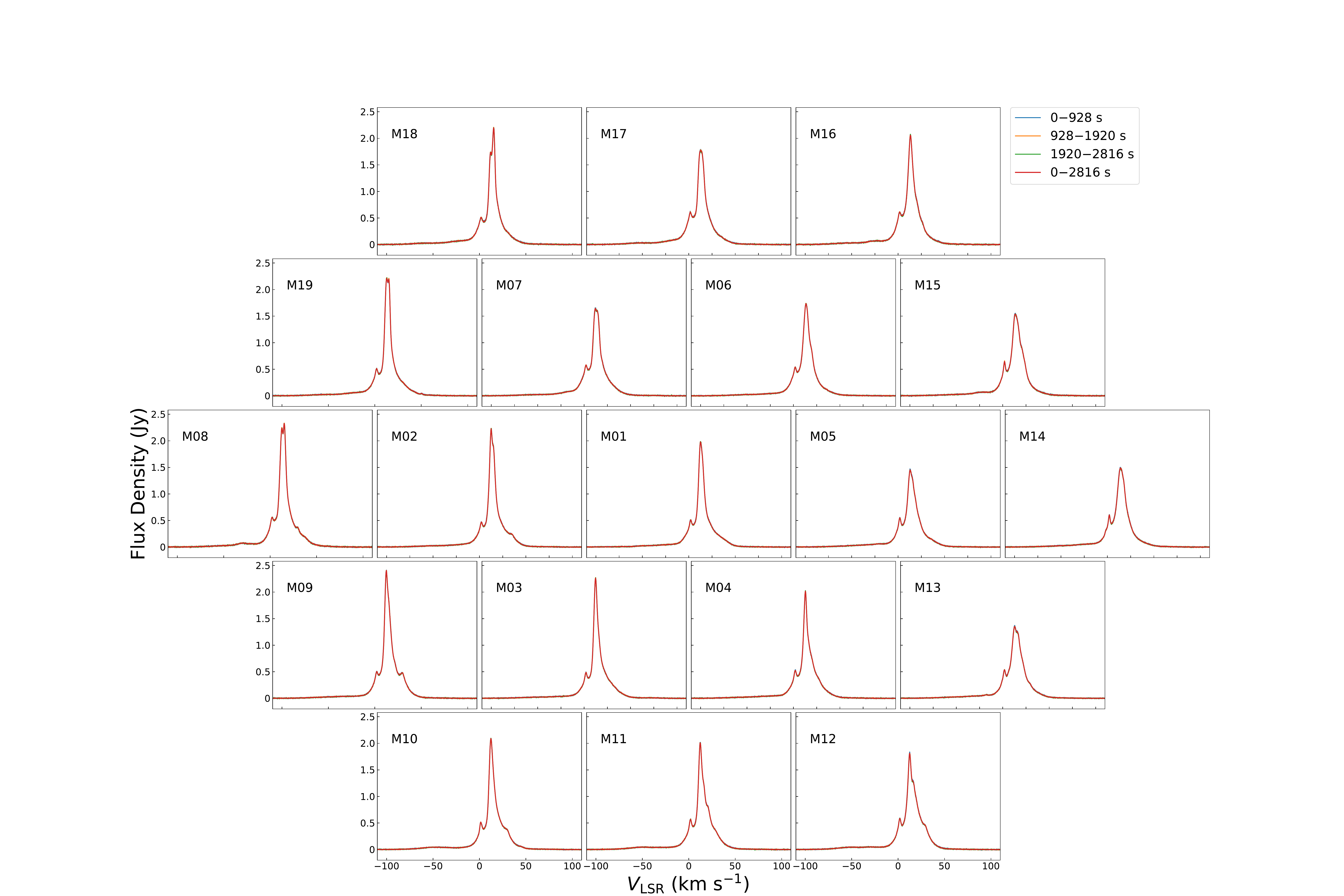}{0.8\textwidth}{(b)}
          }
\caption{(a) The layout of the 19 beams (M01--M09) with the central beam
(M01) pointed to IC\,4997. The color represents
the integrated flux density of the interstellar 21\,cm emission feature.
(b) The spectra obtained with each beam. The scales of all panel are the same. Different colors represent the spectra integrated
over different time spans, which actually superpose each other. \label{fig:bg}}
\end{figure*}

The observations were performed using the tracking mode of 
the FAST 19-beam receiver (see the upper panel of Figure~\ref{fig:bg}) on August 10, 2021. 
The half-power beamwidth (HPBW) of the center beam is about 2.82$\arcmin$ at 1420 MHz. The HPBWs of the outer beams is 
slightly larger than that of the central beam, but the deviation is not more than 0.2$\arcmin$. The standard deviation of pointing accuracy is within 7.9$\arcsec$ \citep{2020RAA....20...64J}.
The central beam (M01) was pointed at the target. As IC\,4997 has a size smaller than the beam size, the other beams sample the adjacent off-source positions. The backend of the spec (F) was used to record the spectral line. This mode records the full frequency range (1.05--1.45\,GHz) of the L-band observation of FAST with 1048576 channels, giving a frequency resolution of 476.84\,Hz or a velocity resolution of 0.10\,km\,s$^{-1}$. A high intensity noise of about 12\,K was injected periodically during about 50 minutes observation times for the flux calibration. The integration time was 3000\,seconds.

The frequency-time `waterfall' plots were examined to remove the radio interference that occasionally appears in the spectra recorded
for all the beam. The  integrated spectra were then obtained for each beam, as shown in the lower panel of Figure~\ref{fig:bg}. 
Considering the potential ripple effect, we fitted the baselines using a sinusoidal function, although a linear fitting does not make any practical difference.
The interstellar 21\,cm emission in the spectra has at least two discrete velocity components, which were not resolved in the observations performed by A86 due to their much poorer spectral resolution.
We integrated the spectra over three different time spans. 
For each beam, there is practically no difference in the different integrated 
spectra, demonstrating the stability of the observations.

In order to investigate the spatial distributions of 
interstellar 21\,cm emission, we derived the flux density
integrated over the velocity range from $-$50\,km\,s$^{-1}$ to 80\,km\,s$^{-1}$
in the local standard of rest (LSR) for each beam. 
The upper panel of Figure~\ref{fig:bg} shows that the spatial variation of
the interstellar 21\,cm line radiation is smooth around IC\,4997,
providing the potential to reliably subtract the interstellar contamination from the on-source spectrum.
The average of the spectra obtained with the six beams (M02--M07) surrounding the central beams was taken as the off-source spectrum.
Considering the potential ripple effect, we fit the the baseline of the `ON minus OFF' spectrum using a sinusoidal function although a linear fitting does not make any practical difference.
The `ON minus OFF' spectrum is shown in the upper panel of Figure~\ref{fig:spc}, which has been binned over ten adjacent frequency channels to improve the signal-to-noise ratio, resulting
in a spectral resolution of 1\,km\,s$^{-1}$.

\section{Results} \label{sec:results}

\begin{figure}[ht!]
\plotone{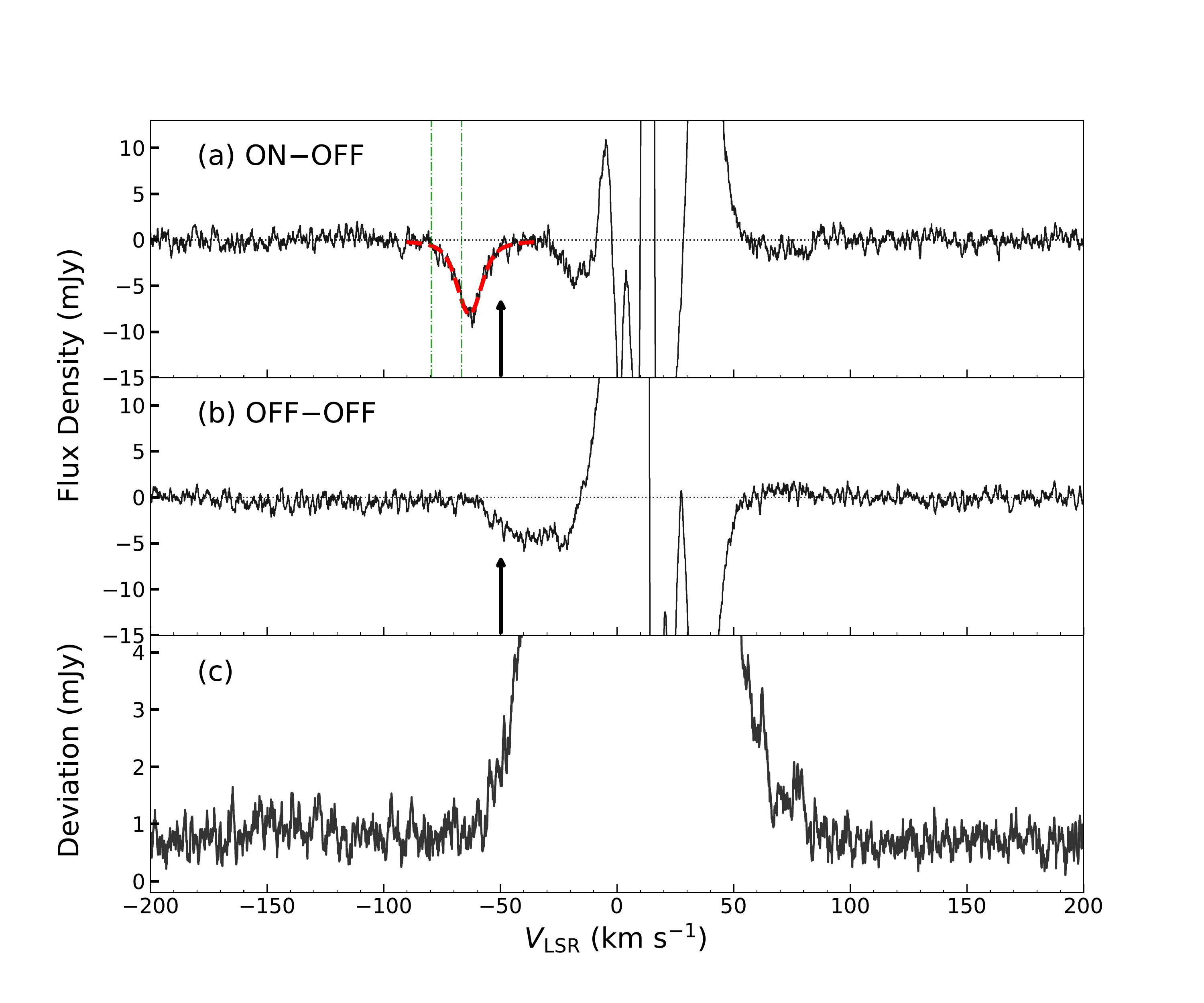}
\caption{(a) The ON spectrum (obtained with M01) subtracted with the OFF spectrum
(the average of the spectra obtained with M02--M07). Note that the baseline has been subtracted.
The red dashed curve represents a Gaussian fitting of the circumstellar \ion{H}{1}  absorption. The arrow marks the $V_{\rm LSR}$ of this PN.
The vertical dashed-dot lines mark the $V_{\rm LSR}$ of
the two components of \ion{Na}{1} D lines
\citep{2020PASP..132g4201R}.
(b) The west-east OFF spectrum (the average of the spectra obtained with
M02 and M05) subtracted with the north-south OFF spectrum
(the average of the spectra obtained with M03, M04, M06, and M07).
 (c) The  deviation of the { individual spectra obtained with M02--M07
with respect to their average (i.e. the OFF spectrum)}.
 \label{fig:spc}}
\end{figure}

\begin{figure}[ht!]
\plotone{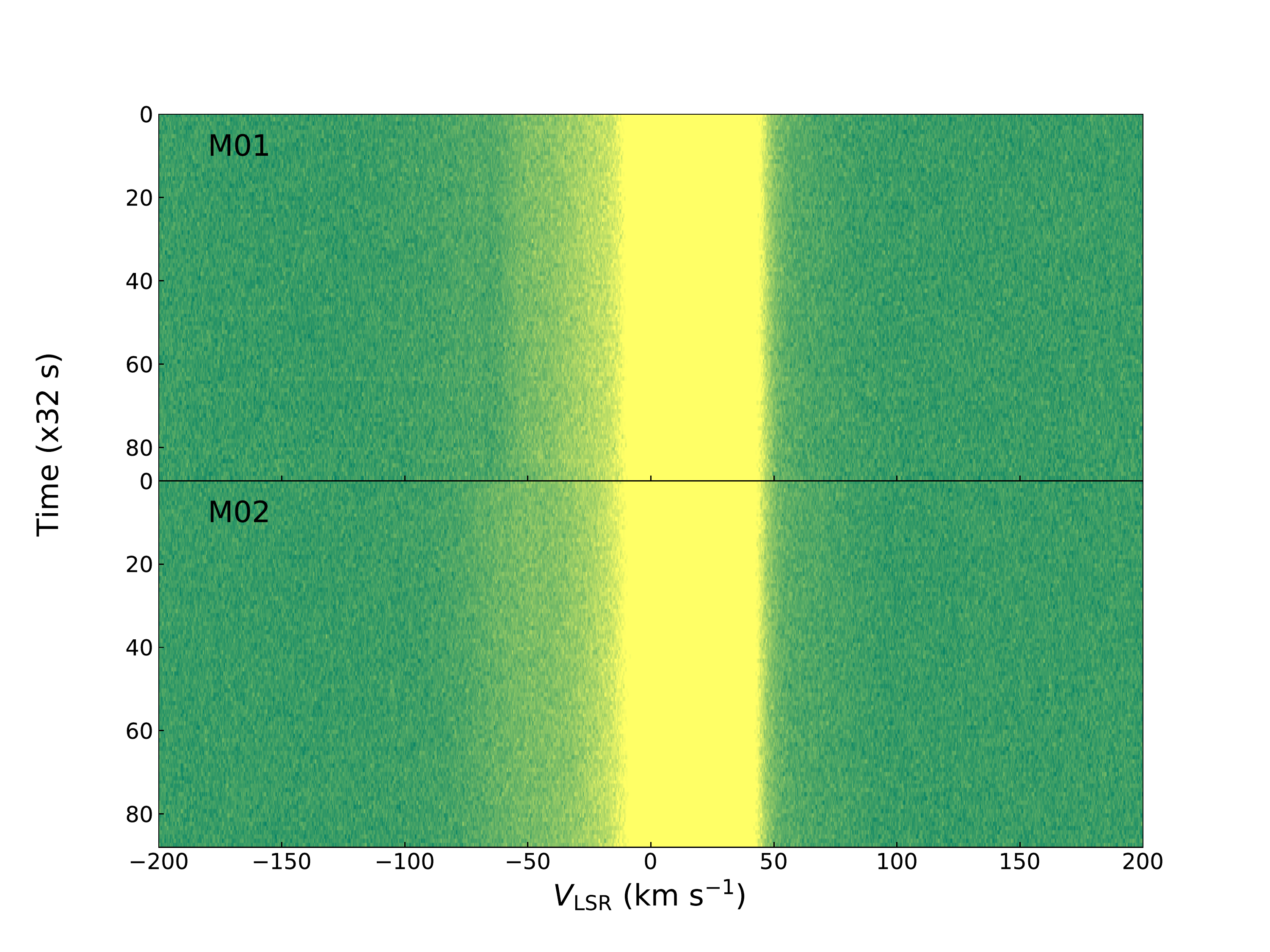}
\caption{The time-velocity waterfall plots 
for beams M01 (upper panel) and M02 (lower panel). 
Note that the absorption feature at $-63.27$\,km\,s$^{-1}$ is visible only in the plot for beam M01.
 \label{fig:falls}}
\end{figure}

IC\,4997 has a $V_{\rm LSR}$ value of  $-49.8\pm1.3$\,km\,s$^{-1}$ 
\citep{1983ApJS...52..399S}. This velocity is marked  in
the upper panel of Figure~\ref{fig:spc}, which clearly reveals a dip just blueward.
We do not see a circumstellar \ion{H}{1} emission feature, which if present should have a LSR velocity of
$-49.8$\,km\,s$^{-1}$.
The subtraction of the off-source spectrum 
is imperfect redward of the target velocity 
because of the strong interstellar emission. 
After closely examining the waterfall plots (Figure~\ref{fig:falls}), we confirm that the dip is stably 
present in the data recorded by beam M01, and is not detected in the other beams.
To further investigate the circumstellar origin of the dip, we derive two OFF-source spectra by averaging the spectra obtained for the west and east beams (M02 and M05)
and those for the north and south beams (M03, M04, M06, and M07), respectively.
The difference between the two OFF-source spectra is shown in the middle panel
of Figure~\ref{fig:spc}, where the dip is invisible and the baseline nearby the
dip is flat.
{ We compute the  deviation ($\sigma_v$) of the individual spectra obtained with beams M02--M07
with respect to their average (i.e. the off-source spectrum described in Section~\ref{sec:ob}), as shown in the  
lower panel of Figure~\ref{fig:spc}. 
$\sigma_v$ has a value of $\sim0.8$\,mJy at the velocity of the dip, 
which}
is much smaller than the peak depth of the dip, and can be taken as the flux uncertainty  introduced by the subtraction of OFF-source spectrum. We thus unambiguously confirm the conclusion of A86 that the dip originates from the 21\,cm absorption on the near side of the expanding \ion{H}{1} envelope around IC\,4997. 
The $\sigma_v$ value however is
 relatively large at $-50<V_{\rm LSR}<80$\,km\,s$^{-1}$ so that 
 features at this velocity range shown
 in the upper panel of Figure~\ref{fig:spc}
are more likely to  be spurious. { 
We can see $\sigma_v\sim2$\,mJy at $v=-49.8$\,km\,s$^{-1}$.
This in part explains the nondetection
of the circumstellar \ion{H}{1} emission feature, whose peak flux density
has an upper limit of $3\sigma_v=6$\,mJy.}

The high-resolution and high-sensitivity FAST spectrum allows us to precisely estimate
the parameters of the circumstellar 21\,cm absorption. By fitting a Voigt function 
to the absorption feature, we obtain the LSR velocity of the peak absorption
$V_{\rm HI}=-63.27\pm0.07$\,km\,s$^{-1}$, 
the full width at half maximum (FWHM) of $12.74\pm0.11$\,km\,s$^{-1}$,
the maximum absorption flux density of  $-8.07\pm0.07$\,mJy,
and the integrated flux density of $-136.7\pm1.5$\,mJy\,km\,s$^{-1}$.
As pointed out by A86, such a $V_{\rm HI}$ value together with the fact that
IC\,4997 is in the first quadrant of the Galactic plane implies that 
the absorption is unlikely to arise from a foreground cloud situated along the line of sight, because otherwise the
distance to this PN or the velocity of the cloud would be unrealistic high.
Assuming the continuum flux density at 21\,cm to be $37\pm2$\,mJy 
\citep{1984AA...130..151I}, we deduce the integration of the optical depth over the velocity $\int\tau dv=3.70\pm0.06$\,km\,s$^{-1}$, which is about 1.8 times higher than that derived by A86.

 The measurement of the FWHM places a loose upper limit on the gas temperature as $T<3600$\,K. 
Given the relatively low critical density ($<1$\,cm\,$^{-3}$)
for the \ion{H}{1} transition, $T$ is expected to be close to
the spin-excitation temperature ($T_{\rm ex}$).
As discussed by A86, $T_{\rm ex}$ is unlikely to be smaller than 50\,K.
To facilitate the comparison of results, we follow
A86 and simply adopt $T \equiv T_{\rm ex}=100$\,K in the following calculations.
{ After deconvolution for the thermal broadening, the \ion{H}{1} line width caused by nebular dynamics is determined to be 
$\Delta V_{\rm HI}=({\rm FWHM}^2-5.5kT/m_{\rm H})^{0.5}=12.6$\,km\,s$^{-1}$, where $m_{\rm H}$ is the mass of the hydrogen atom, and $k$ is the Boltzmann constant. 
If ignoring the projection effect (i.e. assuming a plane-parallel slab) and turbulent broadening and assuming that the velocity increases with radius, 
$\Delta V_{\rm HI}$ represents the velocity difference between the inner and outer \ion{H}{1} layers.}
The  maxima expansion velocity of the \ion{H}{1} shell is approximately determined through
$V_{\rm exp}({\rm H I})=V_{\rm LSR}-V_{\rm HI}+\Delta V_{\rm HI}/2=19.8$\,km\,s$^{-1}$,
which is  higher than the expansion velocity of the inner ionized region \citep[$\sim14.5$\,km\,s$^{-1}$;][]{1984AAS...58..273S}. 
{ If assuming  $T=1000$\,K, $\Delta V_{\rm HI}$ and   $V_{\rm exp}({\rm H I})$ would be scaled down by a factor 
of 1.16 and 1.05, respectively. }
\citet{2020PASP..132g4201R} compare the heliocentric radial velocities
of several atomic and molecular lines, 
(see their Figure 6). For the sake of comparison,
we convert $V_{\rm HI}$
into the heliocentric radial velocity 
and obtain $-80.70$\,km\,s$^{-1}$, which is intermediate between
the heliocentric radial velocities of
[\ion{Fe}{2}] and  H$_2$ emission lines. The \ion{Na}{1} D lines show two absorption
components at $-84$ and $-97$\,km\,s$^{-1}$ with the former stronger than
the latter \citep{2020PASP..132g4201R}.  Our \ion{H}{1} measurement coincides
 with the velocity of the stronger Na D component, while
the weaker velocity component is not detected in the \ion{H}{1} absorption spectrum. { From the report of \citet{2020PASP..132g4201R}, 
the stronger Na D component appears to have a width of 
$\Delta V_{\rm NaI}\sim8$\,km\,s$^{-1}$.  
Considering the projection effect  (see Figure~\ref{fig:str} and the next section), $\Delta V_{\rm NaI}$ may
more accurately represent the velocity difference between the inner and outer neutral layer compared to $\Delta V_{\rm HI}$.
}

Following the approach of A86, we
estimate the mass of circumstellar atomic hydrogen by
\begin{equation}\label{m1}
M_{\rm HI} = 2.14\times10^{-6}\, \Gamma \left( \frac{T_{\rm ex}}{\rm K} \right)\left( \frac{D}{\rm kpc} \right)^2\left( \frac{\theta}{''} \right)^2\left( \frac{\int \tau d\nu}{\rm km\,s^{-1}} \right)\,{\rm M}_\sun,
\end{equation}
where $\Gamma$ is a geometric factor to calibrate the total mass of \ion{H}{1} in the envelope relative to the gas detected in absorption,
$D$ is the distance to IC\,4997, and  $\theta$ is the angular
radius of the \ion{H}{1} envelope.
Note that the mass of atomic hydrogen at the far side of the nebular
is already included in this equation through a spherical symmetry assumption. If the atomic hydrogen is located within a thin shell, we could assume $\Gamma\approx1$,  or otherwise $M_{\rm HI}$ would be underestimated. 


The uncertainty is dominated by the distance measurement. Various distances to IC\,4997 have been reported in the literature. Recently, 
\citet{2016MNRAS.455.1459F} report a value of 4.85\,kpc from their statistical distance estimator based on a robust surface brightness-radius relation, which is more than twice as large as that used in A86. An even larger distance
of 7.12$^{+3.83}_{-1.85}$\,kpc is reported by \citet{2021AA...656A.110C}
based on the Gaia's Early Data Release 3 \citep{2021A&A...649A...1G}. 
We thus adopt $D=5$\,kpc, as a compromise between these recently reported
distances and given the large errors. The angular radius $\theta$ is assumed to be 0.86$\arcsec$ \citep{2021MNRAS.503.2887B}, which is only slightly larger than the value used in A86 (0.8$\arcsec$). Thus we estimate the radius of the atomic hydrogen envelope to be $r({\rm HI})=0.02$\,pc, which indicates a dynamic
age of $t({\rm HI})=r({\rm HI})/V_{\rm exp}({\rm HI})=990$\,yr.

From the above we obtain $M_{\rm HI}=1.46\times10^{-2}$\,M$_\sun$. This is nearly 7.7 times higher than that
derived by A86. Even after correcting for the effects of different distances and $\theta$ values used by us and A86, our calculations still give a \ion{H}{1} mass that is 1.8 times higher. Using the formula given by \citet{1987ApJ...314..572S} but assuming the same $D$ and $\theta$ values as above, we derive the mass of ionized hydrogen
$M_{\rm HII}=2.46\times 10^{-2}$\,M$_\sun$. As a result, the  mass
ratio between atomic and ionized gas ($M_{\rm HI}/M_{\rm HII}$) is about 0.6, suggesting that atomic hydrogen has a significant contribution on nebular mass. Note that the $M_{\rm HI}/M_{\rm HII}$ ratio is distance independent.

{ Under the optically thin assumption, the \ion{H}{1} emission line
can be used to determine $M_{\rm HI}$ by
\begin{equation}\label{m2}
M_{\rm HI} = 2.36\times10^{-4}\left( \frac{D}{\rm kpc} \right)^2\left( \frac{\int F_\nu dv}{\rm mJy\,km\,s^{-1}} \right){\rm M}_\sun,
\end{equation}
where $\int F_\nu dv$ is the integrated flux density of the 
\ion{H}{1} emission line. Assuming that the \ion{H}{1} emission line
has a Gaussian profile with the same FWHM of the absorption line
and based on the measured upper limit of the peak flux density,
we have $\int F_\nu dv < 81.5$\,mJy\,km\,s$^{-1}$, resulting in
$M_{\rm HI}<0.48$\,${\rm M}_\sun$. This is fully compatible with
that obtained using the absorption line. Comparing this with
Equations~\ref{m1}  shows $T_{\rm ex}<1950$\,K.
}


If assuming that the atomic hydrogen is primarily injected from the central star, we can estimate the mass loss rate ($\dot{M}$).
As observations clearly reveal an outwardly accelerating stellar wind
\citep{2020PASP..132g4201R},  we simply assume
a constant acceleration $v\propto t$.  The dynamic age $t({\rm HI})$ then is half
of the age of the \ion{H}{1} shell.
Taking $V_{\rm exp}$ as the velocity of the outer \ion{H}{1} layer,
we roughly estimate the duration time of \ion{H}{1} injection
to be $\Delta t({\rm HI})=2t({\rm HI})\Delta V_{\rm NaI}/V_{\rm exp}=800$\,yr.
{ The thickness of the \ion{H}{1} shell is estimated to be
$\Delta r({\rm HI})=r({\rm HI})[4t^2({\rm HI})-(2t({\rm HI})-\Delta t({\rm HI}))^2]/4t^2({\rm HI})=0.01
$\,pc.
}
Consequently, the mass loss rate of $\dot{M}_{\rm HI} = M_{\rm HI}/\Delta t({\rm HI})$
is derived to be $2\times10^{-5}$\,M$_\sun$ yr$^{-1}$. 
After correcting for the helium abundance \citep{2022MNRAS.510.5984R},
we obtain $\dot{M}=3\times10^{-5}$\,M$_\sun$ yr$^{-1}$.
It should be noted that the value is accurate only to an order of magnitude
since the nebular dynamics are presumably more complex than is assumed here.
The derived $\dot{M}$ is a typical 
value for a PN during the `superwind' epoch at the tip of the AGB \citep{2019NatAs...3..408D}. 
If the stellar ejecta are largely molecular, which  are subsequently dissociated by UV irradiation to enhance 
the abundance of circumstellar \ion{H}{1}, the obtained $\dot{M}$ value
is still valid, but the duration time is that of
injecting the \ion{H}{1} precursor (H$_2$). 
This possibility will be discussed in the following section.

\section{Discussion} \label{sec:discussion}

The total number of H atoms in the atomic shell is $1.7 \times 10^{55}$. Under the thin-shell approximation at $r({\rm HI}) =0.02$\,pc this gives a column density of $N_{\rm HI} = 7.1\times 10^{20}$\,cm$^{-2}$.  \ion{H}{1} can
be formed by photodissociation of the H$_2$ molecules in the original stellar wind, by photons in the 912--1100\AA\ range. The dissociation is limited by self shielding of H$_2$ where the absorption lines become optically thick, and by dust shielding where the far-UV photons suffer heavy extinction. 
Using the formula provided by \citet{2014ApJ...790...10S} and
assuming a solar metallicity, a gas density of $n=100$\,cm\,$^{-3}$, a
standard interstellar radiation field
\citep{1978ApJS...36..595D}, we find that the self shielding 
of H$_2$ yields a $N_{\rm HI}$ value of $6.0\times10^{20}$\,cm$^{-2}$. However, the radiation field near a hot central star of a PN is considerably stronger than the standard interstellar radiation field. The column density to the H$_2$ dissociation front scales as $\chi/n_{\rm HI}$ where $\chi$ is the far-UV flux relative to the standard field. For a stellar blackbody temperature of $T_{\rm eff} = 55,000$\,K \citep{2022A&A...657L...9M} and an assumed luminosity of $7 \times 10^3$\,$L_\odot$, $\chi$ is
of order $5 \times 10^3$. To obtain the measured 
$N_{\rm HI}$ requires a density of $n_{\rm HI} \sim 6 \times 10^5$\,cm$^{-3}$. For comparison, the electron density of the ionized region varies from $n_{\rm e}\sim3\times10^4$ to $\sim10^6$\,cm$^{-3}$
\citep{1994ApJS...93..465H,2022MNRAS.510.5984R}, 
encompassing the derived density of the \ion{H}{1} region.  The \ion{H}{1} shell thickness is roughly $\delta r({\rm HI})\approx N_{\rm HI}/n_{\rm HI}=4\times10^{-4}$\,pc, much lower than $r(\rm HI)$, thus justifying the thin-shell approximation. Note that here we have assumed a molecular instead of an atomic wind as in Section~\ref{sec:results}, and $\delta r({\rm HI})$ represents the mean free path of UV photons. Thus it is not surprising to
see $\delta r({\rm HI}) < \Delta r({\rm HI})$.

Dust shielding becomes important for an extinction of $A_V>2$\,mag. 
The Balmer decrement suggests that the ionized region has an extinction of
$A_V\sim 0.7$\,mag \citep{2010AstL...36..752B}.
Under the assumption of typical interstellar dust, the relation between  $A_V$ and hydrogen column density of \citet{2009MNRAS.400.2050G} suggests that the \ion{H}{1} shell contributes approximately $A_V =0.14$\,mag to the extinction.
However, IC\,4997 is a carbon-rich PN with an unusually high dust-to-gas ratio \citep{1989ApJ...345..306L}.
Therefore, the circumstellar dust may contribute a higher extinction than expected for interstellar silicates. Nevertheless, the agreement with the density of the ionized region suggests that the size of the atomic shell is set by self shielding of H$_2$, and that internal extinction plays only a minor role.

Based on this calculation, we predict that under the thin-shell approximation, the atomic layer surrounding a PN will have a $N_{\rm HI}$ which depends mainly on the stellar temperature and luminosity, and on the inner radius and density of the neutral shell:
\begin{equation}
    N_{\rm HI} \sim 4.4\times10^{14}\,{L_\ast T_{\rm eff} \over r^2_{\rm in} n_{\rm HI} }\,{\rm cm^{-2}} 
\end{equation}
\noindent where $L_\ast$ is the stellar luminosity in solar units, $T_{\rm eff}$ is the effective stellar (blackbody) temperature in K, $r_{\rm in}$ is the inner radius of the atomic shell in units of pc, and $n_{\rm HI} $ is the \ion{H}{1} density in units of cm$^{-3}$.

The photodissociation rate is $k \sim 4 \times 10^{-11} \chi \,\rm s^{-1}$. For $\chi = 5000$, full dissociation therefore takes around 100 years.  This is less than the age of the shell, noting that the star will have been hot enough for dissociation to occur during only part of the time since the ejection of the shell. The \ion{H}{1} shell is therefore likely in equilibrium with the current radiation field. The H$_2$ dissociation front is defined by where the dissociation rate (after self-shielding) equals the H$_2$ formation rate which for our density and assuming normal dust is $\sim 10^{-11}\,\rm s^{-1}$.

The agreement between the derived density of the atomic shell and the ionized region, and that between the \ion{H}{1} expansion velocity and the velocity field of the ionized region indicate that the \ion{H}{1} shell is closely related to the ionized region: it is a continuous shell encompassing an ionization front and a dissociation front.

\begin{figure}[ht!]
\plotone{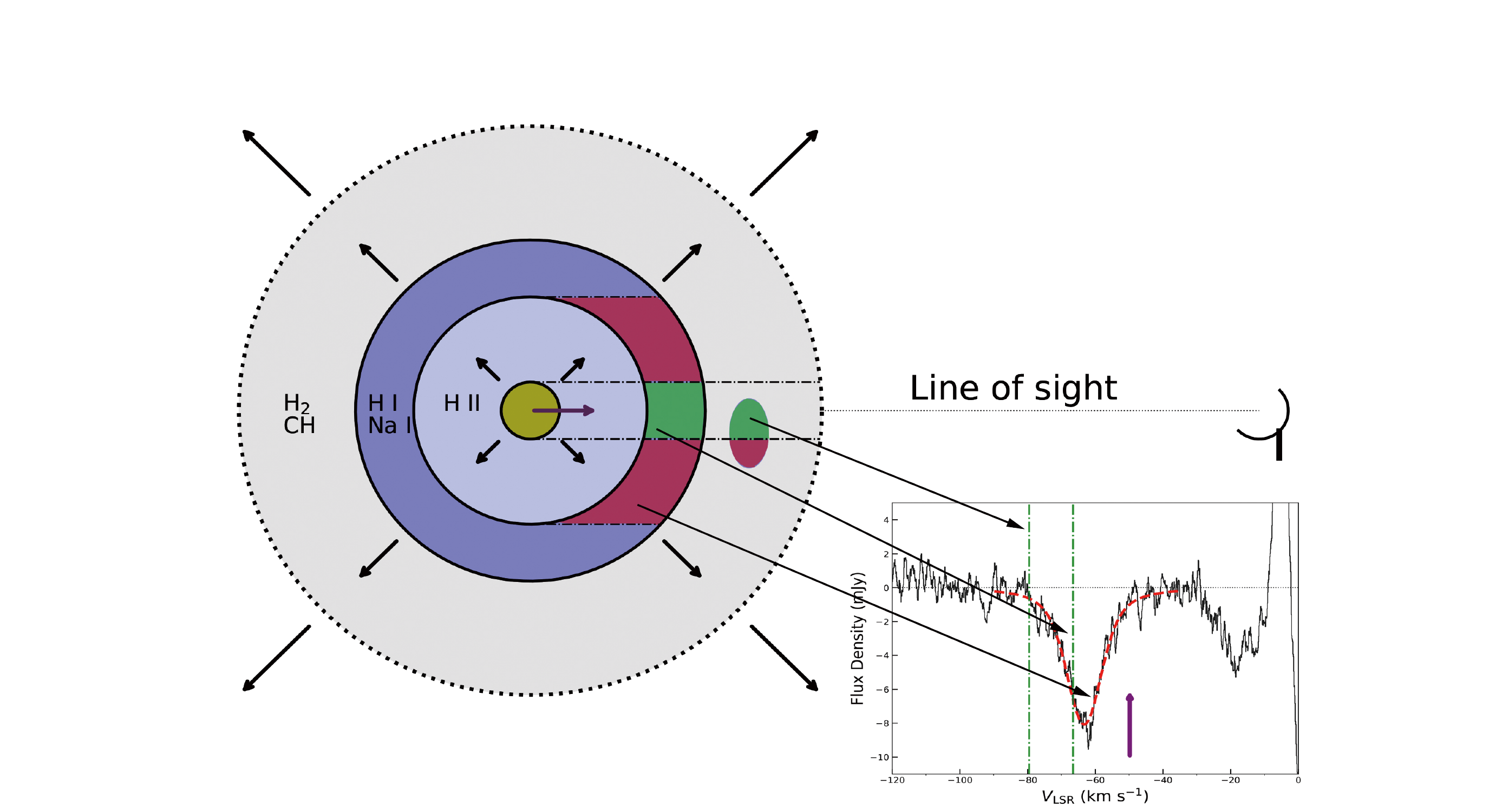}
\caption{The schematic sketch (not drawn to scale) of the nebular structure. 
The central filled circle represents the central
star, which is surrounded from the inside out by
ionized, neutral, and molecular shells. 
A neutral clump lies within the molecular shell.
The radial arrows indicate the outwardly accelerating expansion. The purple arrows represent
$V_{\rm LSR}$.
The regions where the \ion{H}{1} and \ion{Na}{1} absorption features, as shown in Figure~\ref{fig:spc}(a), are formed,  are denoted with red and green, respectively. 
 \label{fig:str}}
\end{figure}

IC\,4997 is a very bright, well studied PN that is characterized by variability in magnitude, emission line ratios, electron temperature and density, line profile, and radio continuum flux density over a timescale of several decades \citep[see][and  references therein]{2022A&A...657L...9M}. 
The variable physical conditions may have led to variable abundance of \ion{H}{1}.  The difference between our results and those of A86 may be due to a  substantial increase of the \ion{H}{1} abundance  during the past 36 years. An increasing number of UV photons (larger $\chi$) can increase the \ion{H}{1} column density over that time period, based on the  photodissociation time scale of H$_2$ calculated above,

In contrast, 
the equivalent widths of the \ion{Na}{1} D absorption lines have decreased by more than half between 1989 and 2014 \citep{2020PASP..132g4201R}, suggesting a more complex picture.
 \citet{2020PASP..132g4201R} show that the \ion{Na}{1} D lines trace two shells, one coincident in velocity with \ion{H}{1} and the other with  a higher expansion velocity. This second shell has a slightly weaker \ion{Na}{1} D absorption depth, and is visible in CH as well. These optical lines are seen in absorption against the central star and therefore trace a smaller region than the radio data which is seen in absorption against the ionized nebula. Comparing the \ion{Na}{1} D and \ion{H}{1} absorption features, the Na column density derived for the velocity component of \ion{H}{1}
is  $N_{\rm NaI}=9.4 \times 10^{11}\,$cm$^{-2}$. This value is not inconsistent with that obtained by \citet{1994ApJ...436..152W}. The faster \ion{Na}{1} D component,
with $N_{\rm NaI}=4.7\times 10^{11}$\,cm$^{-2}$, has no \ion{H}{1} counterpart. 
This discrepancy is most easily explained if this component is not a full shell but a clump in front of the star which does not cover most of the radio-emitting nebula.
The proposed nebular structure is sketched in  Figure~\ref{fig:str},
where the neutral clump and shell 
generate  two \ion{Na}{1} D absorption components of
comparable intensity, while the \ion{H}{1} absorption 
dominantly originates from the shell. It is possible that there  exist more neutral clumps that do not manifest themselves
because of the absence of background continuum emission. 
{ As shown in Figure~\ref{fig:str}, the projection effect of the expanding shell may significantly contribute to the \ion{H}{1} line broadening, while the \ion{Na}{1} D line profile is essentially free from the projection effect. If the radial velocity of the neutral shell is constant and thus the line width is only due to the projection effect, the stronger \ion{Na}{1} D component will peak at the blue edge of the \ion{H}{1} line. This is clearly inconsistent with the observation, and validates again the radial acceleration model.}

{
The estimation of $M_{\rm HI}$ is significantly affected by the uncertainty of $T_{\rm ex}$, as one can tell from Equation~\ref{m1}.
Based on photoionization modelling using Cloudy \citep{2017RMxAA..53..385F}, we find that the \ion{H}{1} region
has an electron temperature of 500 and 300\,K under the assumption
of $n=10^5$ and $4\times10^5$\,cm$^{-3}$, respectively.
This suggests that $T_{\rm ex}$ is most likely in the range 
from 100--500\,K.
If taking $T_{\rm ex}=500$\,K, we would obtain
$M_{\rm HI}=7.3 \times 10^{-2}$\,M$_\sun$. 
Therefore, the \ion{H}{1} absorption measurement indicates that the atomic shell accounts for about 37--75\%\ of the mass of the known nebula (ionized plus atomic). It is a significant nebular component.}
However, the total mass of $M_{\rm HII+HI}=$(3.9--9.8)$ \times 10^{-2}$\,M$_\sun$ is still very low compared to the total mass expected to be lost in a superwind. We suggest that the second component seen in \ion{Na}{1} D is due to a clump in the wind, and does not add significant mass. 
Although the thickness of the \ion{H}{1} shell set by  H$_2$ self shielding
suggests the presence of molecular material outside of this shell,
 the lack of a CO detection does not indicate a  massive molecular envelope. 
\citet{1989ApJ...345..306L} obtained a dust-to-gas ratio of 0.02, yielding
a dust mass of (7.8--19.6)$\times10^{-4}$\,M$_\sun$. Assuming that in reality the dust-to-gas ratio is a more typical 0.01 (limited by the fraction of refractory elements), the undetected molecular mass may contain as much mass as the ionized plus atomic region. That still leaves the total mass of the nebula relatively small.

It follows that, for this object at least, the `missing mass' does not appear to be in the nebula.



\section{Conclusions} \label{sec:con}

Circumstellar \ion{H}{1} absorption is clearly detected in the radio spectrum of IC\,4997, the first such detection from FAST. This feature originates from a shell associated with one of the velocity components of the \ion{Na}{1} D
 lines. Adopting an updated distance of 5\,kpc, we obtain
  $M_{\rm HI}=1.46\times10^{-2}$\,M$_\sun$
  and $N_{\rm HI} = 7.1\times 10^{20}$\,cm$^{-2}$.
  The \ion{H}{1} shell was ejected at least 990\,yr ago during
  the superwind phase.
  The neutral hydrogen in this PN is comparable to the ionized nebula in mass, and may have increased over the past 36 years.
  Nevertheless, the total nebular mass is still far less than
  that expected for the superwind.
 This study demonstrates the feasibility of using FAST to search for atomic hydrogen in a large sample of PNe and
 to provide new insights into the so-called `PN missing mass problem'.

\begin{acknowledgments}

We would like to thank an anonymous referee for helpful comments and
Ning-Yu Tang and Nai-Ping Yu for their assistance with the data reduction.
This work was supported by the National Science Foundation of China (NSFC, Grant No. 11973099) and the science research grants from the China Manned Space Project (NO. CMS-CSST-2021-A09 and CMS-CSST-2021-A10). QP thanks the Hong Kong Research Grants Council for GRF research support under grants 17326116 and 17300417. FAST is a National Major Scientific Project built by the Chinese Academy of Sciences. Funding for the project has been provided by the National Development and Reform Commission. FAST is operated and managed by the National Astronomical Observatories, Chinese Academy of Sciences.

\end{acknowledgments}

\end{document}